\documentclass[10pt,conference]{IEEEtran}
\usepackage{amsmath, amssymb, epsfig}
\usepackage{epsf,citesort}

\newcommand{\oT}{\overline{T}} 
\newcommand{\tx}{\widetilde{x}} 
\newcommand{\hx}{\widehat{x}} 
 
\newcommand{\mN}{\mathcal{N}} 

\newtheorem{theorem}{Theorem}
\newtheorem{lemma}{Lemma}

\newtheorem{prop}{Proposition}

\begin{document}

\title{Equivalence of LP Relaxation and Max-Product for Weighted
  Matching in General Graphs}

\author{
\authorblockN{Sujay Sanghavi}
\authorblockA{LIDS, MIT \\
sanghavi@mit.edu}}

\maketitle

\begin{abstract}

Max-product belief propagation is a local, iterative algorithm to find
the mode/MAP estimate of a probability distribution. While it has been
successfully employed in a wide variety of applications, there are
relatively few theoretical guarantees of convergence and correctness
for general loopy graphs that may have many short cycles. Of these,
even fewer provide exact ``necessary and sufficient''
characterizations.

In this paper we investigate the problem of using max-product to find the
maximum weight matching in an arbitrary graph with edge weights. This
is done by first constructing a probability distribution whose mode
corresponds to the optimal matching, and then running max-product.
Weighted matching can also be posed as an integer program, for which
there is an LP relaxation. This relaxation is not always tight. In
this paper we show that 
\begin{enumerate}
\item If the LP relaxation is tight, then max-product always
  converges, and that too to the correct answer.
\item If the LP relaxation is loose, then max-product does not
  converge.
\end{enumerate}
This provides an exact, data-dependent characterization of max-product
performance, and a precise connection to LP relaxation, which is a
well-studied optimization technique. Also, since LP relaxation is
known to be tight for bipartite graphs, our results generalize other
recent results on using max-product to find weighted matchings in
bipartite graphs.
\end{abstract}

\section{Introduction}

Message-passing algorithms, like Belief Propagation and its variants
and generalizations, have been shown empirically to be very effective
in solving many instances of hard/computationally intensive problems
in a wide range of fields. These algorithms were originally designed
for exact inference (i.e. calculation of marginals/max-marginals) in
tree-structured probability distributions. Their application to
general graphs involves replicating their iterative local update rules
on the general graph. In this case however, there are no guarantees of
either convergence or correctness in general.

Understanding and characterizing the performance of message-passing
algorithms in general graphs remains an active research area.
\cite{bib:aji_single_cycle,bib:weiss_single_cycle} show correctness
for graphs with at most one
cycle. \cite{bib:weiss_freeman_gaussian,bib:walksums} show that for
gaussian problems the sum-product algorithm finds the correct means
upon convergence, but does not always find the correct variances.
\cite{bib:rich_urb,bib:van_roy} show asymptotic correctness for random
graphs associated with decoding.  \cite{bib:weiss_local_opt} shows
that if max-product converges, then it is optimal in a relatively
large ``local'' neighborhood.

In this paper we consider the problem of using max-product to find the
maximum weight matching in an arbitrary graph with arbitrary edge
weights. This problem can be formulated as an integer program, which
has a natural LP relaxation. In this paper we prove the following
\begin{enumerate}
\item If the LP relaxation is tight, then max-product always
  converges, and that too to the correct answer.
\item If the LP relaxation is loose, then max-product does not
  converge.
\end{enumerate}

Bayati, Shah and Sharma \cite{bib:bayati} were the first to
investigate max-product for the weighted matching problem. They showed
that if the graph is bipartite then max-product always converges to
the correct answer. Recently, this result has been extended to
$b$-matchings on bipartite graphs \cite{bib:b_match}. Since the LP
relaxation is always tight for bipartite graphs, the first part of our
results recover their results and can be viewed as the correct
generalization to arbitrary graphs, since in this case the tightness
is a function of structure as well as weights.

We would like to point out three features of our work:
\begin{enumerate}
\item It provides a {\em necessary and sufficient} condition for
  convergnce of max-product in arbitrary problem instances. There are
  very few non-trivial classes of problems for which there is such a
  tight characterization of message-passing performance.
\item The characterization is {\em data dependent}: it is decided
  based not only on the graph structure but also on the weights of the
  particular instance. 
\item Tightness of LP relaxations is well-studied for broad classes of
  problems, making this chracterization promising in terms of both
  understanding and development of new algorithms.
\end{enumerate}
Relations, similarities and comparisons between max-product and linear
programming have been used/mentioned by several authors
\cite{bib:yanover_lp_bp,bib:wainw_linprog,bib:feldman_allerton}, and
an exact characterization of this relationship in general remains an
interesting endeavor. In particular, it would be interesting to
investigate the implications of these results as regards elucidating
the relationship between iterative decoding of channel codes and LP
decoding \cite{bib:lp_decoding}.

\section{Weighted Matching and its LP Relaxation}

A {\em matching} in a graph is a set of edges such that no two edges
in the set are incident on the same node. Given a graph $G = (V,E)$,
with non-negative weights $w_e$ on the edges $e\in E$, the {\em
weighted matching problem} is to find the matching $M^*$ whose edges
have the highest total weight. In this paper we find it convenient to
refer to edges both as $e\in E$ and as $(i,j)$, where $i,j \in V$.

Weighted matching can be written as the following integer program
(IP):
\[
\max \quad \sum w_e x_e
\]
\begin{equation}\label{eq:ip}
s.t. \sum_{j\in \mN(i)} x_{ij} \leq 1 \quad \text{for all $i\in V$}
\end{equation}
\[
x_e \in \{0,1\} \quad \text{for all $e\in E$}
\]

The LP relaxation of the above problem is to replace the constraint
$x_e \in \{0,1\}$ with the constraint $x_e \geq 0$. {\em This
relaxation is in general not tight}, i.e. there might exist
non-integer solutions with strictly higher value than any integral
solution. It is known however that the LP relaxation is {\em always}
tight for bipartite graphs: no matter what the edge weights, the
bipartite-ness ensures tightness of the LP relaxation. If a graph is
not bipartite, the tightness of the LP relaxation will depend on the
edge weights: the same graph may have tightness for one set of weights
and looseness for another set.  

The dual of the above linear program is the {\em vertex cover}
problem: minimize the total of the weights $z_i$ that need to be
placed on nodes so as to ``cover'' the edge weights: (DP)
\[
\min \quad \sum z_i 
\]
\[
s.t. \quad w_{ij} \leq z_i + z_j \quad \text{for all $(i,j)\in E$}
\]
\[
z_i \geq 0 \quad \text{for all $i$}
\]

\begin{lemma}[complimentary slackness]\label{lem:lp_tight}
When the LP relaxation is tight, the optimal matching $M^*$ and the
optimal dual variables $z$ and satisfy the following properties:
\begin{enumerate}
\item if $(i,j)\in M^*$ then $w_{ij} = z_i + z_j$
\item if $(i,j)\notin M^*$ then $w_{ij} \leq z_i + z_j$
\item if no edge in $M^*$ is incident on node $i$, then $z_i = 0$
\item $z_i \leq \max_e w_e$ for all $i$
\end{enumerate}
\end{lemma}

\section{Background on the Max-Product Algorithm}\label{sec:mp_back}

The {\em factor graph} \cite{bib:ksh_frey_loel} of a probability distribution
represents the conditional independencies of the distribution.  The
Max-Product (MP) algorithm is a simple, local, iterative message
passing algorithm that can be used (in an attempt) to find the
mode/MAP estimate of a probability distribution. Nodes and factors
pass messages to each other, and nodes maintain ``beliefs'', which
represent the max-marginals.  When max-product is applied to problems
involving general ``loopy'' graphs, one of the following three
scenarios may result:
\begin{enumerate}
\item The algorithm may not converge.
\item The algorithm may converge, but to an incorrect answer.
\item The algorithm may converge to the correct answer.
\end{enumerate}
As has been mentioned, here has been siginifcant work attempting to
understand the properties of MP for loopy graphs. For the results in
this paper, we will use the following two insights:
\begin{enumerate}
\item At any time, the belief of the max-product algorithm for a given
  variable corresponds to the belief at the root of the corresponding
  {\em computation tree} distribution \cite{bib:weiss_single_cycle}
  associated with that variable at that time.  We describe what this
  computation tree distribution corresponds to for the weighted
  matching problem in the next section.
\item If max-product {\em does} converge, the resulting beliefs are
  optimal in a large ``local'' neighborhood
  \cite{bib:weiss_local_opt}: let $\hx$ be the assignment as given by
  the converged max-product and $\tx$ be any other assignment. If the
  variables assigned different values in $\hx$ and $\tx$ form an
  induced graph containing at most one cycle in each component, then
  $p(\hx)\geq p(\tx)$.
\end{enumerate} 

\section{Max-Product for Weighted Matching}

The problem of finding $M^*$ can be formulated as the problem of
finding the mode of a suitably (artifically) constructed probability
distribution $p$. In fact, there are in general {\em several} ways to
construct this distribution for the {\em same} instance of a graph
$G$. We now present one construction\footnote{This construction is
different from the one in \cite{bib:bayati}, which had a pairwise
model with variables corresponding to nodes in the graph. However, the
results of this paper continue to hold when the construciton in
\cite{bib:bayati} is modified to be applicable to general graphs}.

Associate a binary variable $x_e\in \{0,1\}$ with each edge $e\in E$,
and let
\begin{equation}
\label{eq:p}
p(x) ~~ = ~~ \frac{1}{Z} \prod_{i\in V} \mathbf{1}_{\{\sum_{j\in
\mN(i)} x_{ij} \,\leq \,1 \}} ~ \prod_{e\in E} e^{w_e x_e}
\end{equation}
Here $\mN(i)$ represents the neighborhood of node $i$ in $G$, and $Z$
is a normalizing constant. The variable $x_e$ can be interpreted as
follows: $x_e = 1$ indicates that $e\in M^*$, while $x_e = 0$
indicates $e\notin M^*$.  The term $\mathbf{1}_{\{\sum_{j\in \mN(i)}
x_{ij} \leq 1 \}}$ enforces the cosntraint that of the edges incident
to node $i$, at most one can be assigned the value ``1''. Thus, it is
easy to see that $p(x)>0$ if and only if the edges with $x_e =1$
constitute a matching in $G$. Furthermore, the mode of $p$ corresponds
to the max-weight matching $M^*$.

The factor graph max-product involves messages between variables and
factors. In our case the variables are the edges $(i,j)\in E$, and the
factors are nodes $i\in V$. Thus at any time $t$ there will be
messages $m_{i\rightarrow (i,j)}^t$ from node (factor) $i$ to edge
(variable) $(i,j)$, as well as messages $m_{(i,j)\rightarrow i}^t$.
Each message will be a length-two vector of real numbers, indexed by 0
and 1. The message update rules can be simplified to the following:
\begin{eqnarray*}
m_{(i,j)\rightarrow i}^{t+1}[1] & = & e^{w_{ij}} m_{j\rightarrow
  (i,j)}^t[1] \\
m_{(i,j)\rightarrow i}^{t+1}[0] & = & m_{j\rightarrow (i,j)}^t[0] \\
m_{i\rightarrow (i,j)}^{t+1}[1] & = & \prod_{k\in \mN(i)-j}
  m_{(k,i)\rightarrow i}^t [0] \\
m_{i\rightarrow (i,j)}^{t+1}[0] & = & \max {\huge \{} \prod_{k\in
  \mN(i)-j} m_{(k,i)\rightarrow i}^t [0] ~~, 
\\  & & \quad \quad \quad \max_{k\in \mN(i)-j} m_{(k,i)\rightarrow
  i}^t[1] {\huge \} } 
\end{eqnarray*}
Also, at every time each edge (variable) maintains a belief vector
$b_{(i,j)}^t$ as follows:
\begin{eqnarray*}
b_{(i,j)}^t[0] & = & m_{i\rightarrow (i,j)}^t[0] ~ \times ~
m_{j\rightarrow (i,j)}^t[0] \\
b_{(i,j)}^t[1] & = & e^{w_{ij}} m_{i\rightarrow (i,j)}^t[1] ~ \times ~
m_{j\rightarrow (i,j)}^t[1] 
\end{eqnarray*}
The $p$ defined above can be used to find $M^*$ as follows: first run
max-product. At any time $t$ and for each edge $e$ there will be two
beliefs $b_e^t[0]$ and $b_e^t[1]$. If max-product converges, assign to
each variable the value (i.e. ``0'' or ``1'') that corresponds to the
stronger belief. Then, declare the set of all edges set to ``1'' to be
the max-product output.

\subsection{The Computation Tree for Weighted Matching}

Our proofs rely on the computation tree interpretation
\cite{bib:weiss_single_cycle,bib:tatik_jordan} of the Max-product
beliefs. We now describe this interpretation when max-product is
applied to $p$ as given in (\ref{eq:p}).

For an edge $e$ let $\oT_e(k)$ be the {\em full depth-$k$ computation
tree rooted at $e$}. This is generated recursively: take $\oT_e(k-1)$
and to each leaf $v$ add as children a copy of each of the neighbors
of $v$ in $G$, except for the unique neighbor of $v$ which is already
present in $\oT_e(k-1)$. Also, each new edge has the same weight as
its copy in the original $G$. The recursion is started with the
single-edge tree $\oT_e(1) = e$, both of whose endpoints are
leaves. This initial edge is the {\em root} of $\oT_e$.

Consider now the ``full synchronous'' max-product, where at each time
every message in the network is updated. In this case the computation
tree $T_e(k)$ for edge $e$ at time $k$ will be
$\oT_e(k)$. Alternatively, max-product may be executed asynchronously
with only a subset of the messages updated in every time slot. In this
case $T_e(k)$ will be a sub-tree of $\oT_e(k)$. In either case, the
computation tree interpretation states at time $k$ we have
$b^k_e[1]>b^k_e[0]$ if and only if the root of $T_e(k)$ is a member of
a max-weight matching on the tree $T_e(k)$.

The figure below shows an example where on the left is $G$: the
four-cycle $abcd$ and the chord $ac$, with a matching $M =
\{(a,b),(c,d)\}$ depicted in bold. On the right is the computation
tree $\oT_{(a,b)}(4)$ which is the full tree of depth 4 rooted at edge
$(a,b)$. The bold edges depict the projection $M_T$ of $M$ onto
$\oT_{(a,b)}(4)$: an edge $e$ in the tree is in $M_T$ if and only if
its copy in $G$ is in $M$.

\begin{figure}[h]
\centering
\epsfig{file=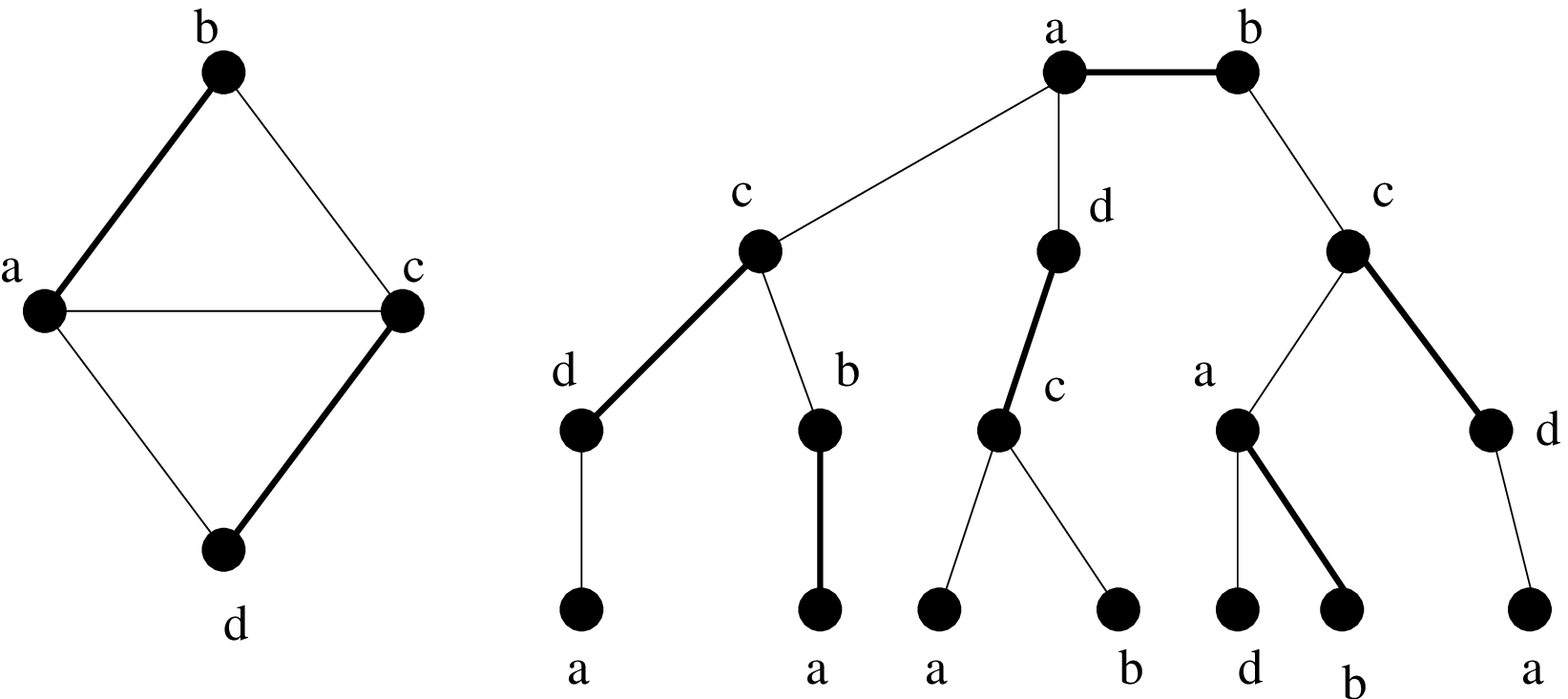,height=1.3in}
\end{figure}

\begin{lemma}\label{lem:matching_corresp}
Let $M$ be a matching in $G$ and $T_e(k)$ be a computation tree. Let
$M_T$ be the set of all copies in $T_e(k)$ of all edges in $M$. Then,
$M_T$ is a matching in $T_e(k)$. Also, if $M$ is maximal in $G$, $M_T$
is maximal in $T_e$.
\end{lemma}

Of course $T_e$ will also contain other matchings that are not
projections of matchings in $G$. Finally, we say that a (possibly not
full) tree $T_e(k)$ is {\em full upto depth $k_1$} if the full tree
$\oT_e(k_1)$ is contained in $T_e(k)$.

\section{Equivalence of Max-Product and LP Relaxation}

We are now ready to prove the main result of this paper: the
equivalence of Max-Product and LP Relaxation. Before we proceed, we
define the following terms
\begin{enumerate}
\item We say that the {\em LP relaxation is tight} if the linear
  program (LP) obtained by relaxing the integer program (\ref{eq:ip})
  has a unique optimal solution at which all values $x_e$ are either 0
  or 1.
\item We say that {\em max-product converges by step $k$} if the
  variable assignments (0 or 1) that maximize the beliefs at each node
  remain constant once the associated computation tree is full up to
  depth at least $k$. Note that this includes both synchronous and
  asynchronous message updates. We say that {\em max-product
  converges} if there exists some $k<\infty$ such that max-product
  converges by step $k$. Finally, we say that {\em max product
  converges to the correct answer} if the beliefs $b_e$ at convergence
  are such that $b_e[1]>b_e[0]$ if and only if $e\in M^*$, and
  $b_e[1]<b_e[0]$ if and only if $e\notin M^*$
\end{enumerate}

We also need to make some uniqueness assumptions. It is
well-recognized that max-product may perform poorly in the presence of
multiple optima, and that characterizing performance in this case is
hard. For the rest of this paper we will assume the following:
\begin{enumerate}
\item[{\bf A1}] $M^*$ is the unique optimal matching.
\item[{\bf A2}] The linear program always has a unique optimal
  solution. Note that this can be fractional, but it has to be unique.
\end{enumerate}

\subsection{Max-product is as Powerful as LP Relaxation}

In this section we prove that if the LP relaxation is tight then
Max-Product converges to the correct answer. Recall that when the LP
is tight, part 2 of Lemma \ref{lem:lp_tight} says that if $(i,j)\notin
M^*$ then $w_{ij} \leq z_i + z_j$. The uniqueness assumptions {\bf
A1-2} further imply that the inequality is strict: $w_{ij} < z_i +
z_j$. Another way of saying this is that there exists an $\epsilon>0$
such that
\begin{equation}\label{eq:gap}
w_{ij} \leq z_i + z_j - \epsilon \quad\text{for all $(i,j)\notin M^*$}
\end{equation}

\begin{theorem}
Consider a weighted graph $G$ for which the LP relaxation is
tight. Then max-product converges to the correct answer by step
$\frac{2w_{max}}{\epsilon}$, where $w_{max} = \max_e w_e$ is the
weight of the heaviest edge, and $\epsilon$ satisfies (\ref{eq:gap}).
\end{theorem}

\noindent{\em Proof:}

Let $M^*$ be the optimal matching on $G$. For max-product to be
convergent and correct, we need that $b^t_e[1] > b^t_e[0]$ for all
$e\in M^*$ and $b^t_e[1] < b^t_e[0]$ for all $e\notin M^*$, and for
all $t$ such that $T_e(t)$ is full upto depth
$\frac{2w_{max}}{\epsilon}$.

So suppose that for such a $t$ there exists an $e\notin M^*$ such that
$b^t_e[1] > b^t_e[0]$. Then, there exists a matching $M$ in $T_e(t)$
such that {\em (a)} the root $e\in M$, and {\em (b)} $M$ has the
largest weight among matchings on $T_e(t)$. Let $M^*_T$ be the set of
all edges in $T_e(t)$ that are copies of edges in $M^*$. By lemma
\ref{lem:matching_corresp}, $M^*_T$ is a maximal matching on
$T_e(t)$. Also, the root $e\notin M^*_T$ by assumption.

The symmetric difference $M^*_T\triangle M$ consists of disjoint
alternating paths in $T_e(t)$: each path will have every alternate
edge in $M^*_T$ and all other edges in $M$. Let $P$ be the path that
contains the root $e$.  We now show that $w(P\cap M^*_T) > w(P\cap
M)$.

Recall that the optimal dual solution assigns to each node $i$ in $G$
a ``dual value'' $z_i \geq 0$. Associate now with each node in $T_e(t)$ the
dual value of its copy in $G$. Then, by Lemma \ref{lem:lp_tight} we
have that $w_{ij} = z_i + z_j$ for each $(i,j)\in P\cap
M^*_T$. Suppose now that neither endpoint of $P$ is a leaf of 
$T_e(t)$.  In this case, we have
\[
w(P\cap M^*_T) ~ = ~  \sum_{i\in P} z_i 
\]
On the other hand, we know that (\ref{eq:gap}) holds for each edge in
$P\cap M$. Adding these up gives
\[
w(P\cap M) ~ \leq ~ \sum_{i\in P} z_i ~ - \epsilon |P\cap M| 
\]
By assumption, the root $e\in P\cap M$, so $|P\cap M|\geq 1$ and hence
$w(P\cap M^*_T) > w(P\cap M)$ when no endpoints of $P$ are leaves.

Suppose now that exactly one of the endpoints $v$ of $P$ is a leaf
of $T_e(t)$. In this case, we have that
\begin{eqnarray*}
w(P\cap M^*_T) & \geq &  \sum_{i\in P} z_i ~ - \,z_v ~ \geq ~
\sum_{i\in P} z_i ~ - \, w_{max} 
\end{eqnarray*}
where the last inequality follows from part 4 of Lemma \ref{lem:lp_tight}
Also, $T_e(t)$ is assumed to be full up to depth $k$, so this implies that
$|P\cap M|\geq \frac{k}{2}$. This means that
\[
w(P\cap M) ~ \leq ~ \sum_{i\in P} z_i ~ - \epsilon \,\frac{k}{2}
\]
Now, since $k\geq \frac{2w_{max}}{\epsilon}$, this implies that
$w(P\cap M^*_T) > w(P\cap M)$. The final case, where both endpoints of
$P$ are leaves, works out in the same way, except that now $|P\cap
M|\geq k$ and $w(P\cap M^*_T) \geq \sum_{i\in P} z_i ~ - \, 2w_{max}$.

Thus, in any case, we have that $w(P\cap M^*_T) > w(P\cap
M)$. Consider now the set of edges $M-(P\cap M) + (P\cap M^*_T)$. This
set forms a matching on $T_e(t)$, and has higher weight than $M$. This
contradicts the choice of $M$, and so establishes that $b^t_e[1] <
b^t_e[0]$ for all $e\notin M^*$. A similar contradiction argument can
be used to establish that $b^t_e[1] > b^t_e[0]$ for all $e\in
M^*$. This completes the proof. \hfill $\blacksquare$

\subsection{LP Relaxation is as Powerful as Max-product}

In this section we prove that if the LP relaxation is loose then
max-product does not converge to the correct answer. Before we do so
however, we note that this implies a stronger result: that when LP is
loose then in fact max-product does not converge at all. 

\begin{lemma}\label{lem:local_global}
Consider the distribution $p(x)$ as given in (\ref{eq:p}). If
Max-Product converges, then its output exactly corresponds to the true
optimal matching $M^*$.
\end{lemma}

The proof of this lemma uses the ``local optimality'' result of Weiss
and Freeman \cite{bib:weiss_local_opt}. In particular, for $p$ it turns
out that local optimality implies global optimality. This means that
it is not possible for max-product to converge to an incorrect answer:
it will either not converge at all, or converge to $M^*$. We do not
use this explicitly in the proofs below, but it strengthens the
results as mentioned above.

We now proceed with showing that max-product does not converge to the
correct $M^*$ when LP is loose. As a first step, we need a
combinatorial characterization of when the LP relaxation is loose. We
now make some definitions. We say that a node $v$ is {\em saturated}
by a matching $M$ if there exists an edge $e\in M$ that is incident to
$v$.

A {\em blossom} with respect to a matching $M$ is an odd cycle $C$
with $\frac{|C|-1}{2}$ edges in $M$.\footnote{Blossoms were first
defined in \cite{bib:edmonds}, which also provided the first efficient
algorithm for weighted matching in arbitrary graphs.} Note that a
blossom has a unique {\em base}: a node not saturated by any edge in
$C\cap M$. A {\em stemmed blossom} $B_1$ (w.r.t $M$) is a blossom $C$,
along with an alternating path (stem) $P$ that starts at the base of
$C$, and starts with an edge in $M$. Also, $P$ should be such that the
set $M-(P\cap M) + (P-M)$ remains a matching in $G$.  

A {\em bad stemmed blossom} is one in which the edge weights satisfy
\[
w(C\cap M) + 2w(P\cap M) ~ < ~ w(C-M) + 2w(P-M)
\]
Note that it may well be the case that $|P|=0$, in which case $B_1$ is
just an odd cycle. The following is an example of a bad stemmed
blossom. The bold edges are the ones in $M$, the numbers denote the
weights of the corresponding edges, and the last node $i$ has no edge
of $M$ incident on it. The blossom $C$ in this case is the cycle
$abcde$, and node $c$ is its base. The path/stem $P$ is $cfghi$.

\begin{figure}[h]
\centering
\epsfig{file=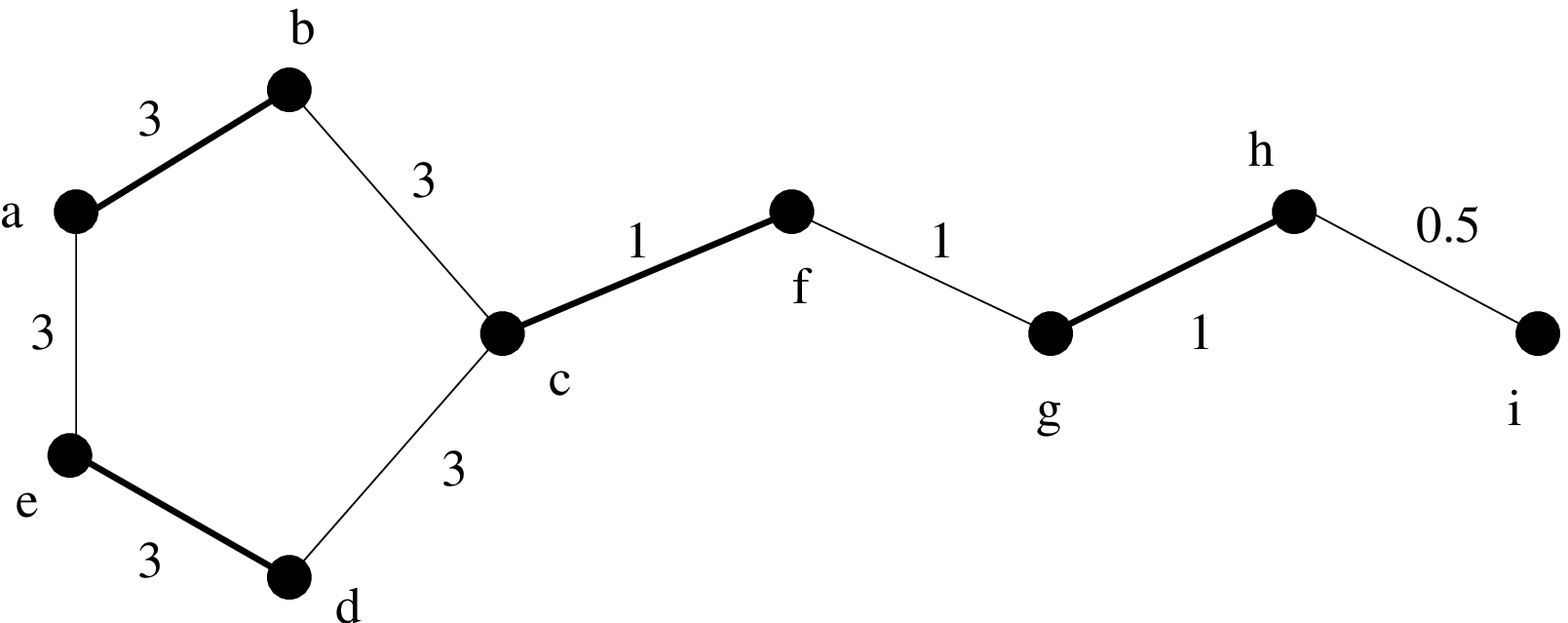,height=1.0in}
\end{figure}

A {\em blossom pair} $B_2$ is two blossoms $C_1$ and $C_2$ and an
alternating path $P$ between the bases of the two blossoms such that
$P$ begins and ends with edges in $M$. A {\em bad blossom pair} is one
in which the edge weights satisfy
\begin{eqnarray*}
& w(C_1\cap M) + w(C_2\cap M) + 2w(P\cap M)  \quad & \\ 
& \quad \quad < ~ w(C_1-M) + w(C_2-M) + 2w(P-M) & 
\end{eqnarray*}
The following is an example of a bad blossom pair.

\begin{figure}[h]
\centering
\epsfig{file=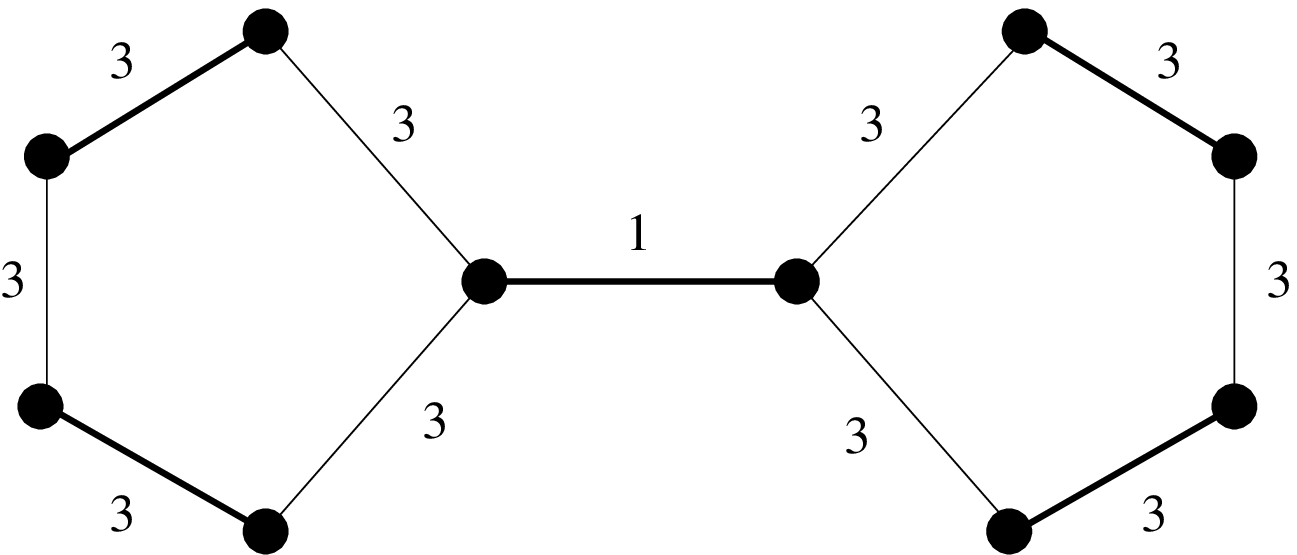,height=1.0in}
\end{figure}

The following proposition provides a combinatorial characterization of
when the LP relaxation is loose, and is crucial to the proof of the
subsequent theorem.

\begin{prop}\label{prop:bad_subgr}
If the LP relaxation is loose, then there exists a bad stemmed
blossom, or a bad blossom pair, with respect to the optimal matching
$M^*$.
\end{prop}

{\em Proof:} In appendix.

We use the presence of these ``bad'' subgraphs in $G$ to show that
max-product does not converge to the correct answer.  Before we do so,
we need one additional lemma. This states that if max-product
converges by step $k$ to some matching $M$ on $G$, then the optimal
matching $M_T$ on the computation tree looks like $M$ in the
neighborhood of the root.

\begin{lemma} \label{lem:convg_tree}
Suppose max-product converges to a matching $M$ in $G$ by step
$k$. Consider any edge $e$, some $m\geq 1$ and a corresponding
computation tree $T_e$ which is full up to depth $k+m$. Let $M_T$ be
the max-weight matching on the tree. Then, for any edge $f\in T_e$
that is within distance $m$ of the root $e$, $f\in M_T$ if and only if
its copy $f_1$ in $G$ is such that $f_1 \in M$.
\end{lemma}

Note that the above lemma also applies to the root $e$ of the tree. We
are now ready to state and prove the main result of this
section. Recall that the belief $b_e$ on an edge at convergence is
incorrect if either $e\in M^*$ but $b_e[0]>b_e[1]$, or $e\notin M^*$
but $b_e[1]>b_e[0]$.

\begin{theorem}
Consider a weighted graph $G$ for which the LP relaxation is
loose. Then, the max-product beliefs do not converge to the correct
$M^*$: for any given $k$, there exists a $k_1 \geq k$ and computation
trees $T_e, e\in E$ such that each $T_e$ is full upto depth $k_1$, but
the beliefs on some of the edges are incorrect. Lemma
\ref{lem:local_global} further implies that in fact in this case
max-product does not converge at all.
\end{theorem}

\noindent {\em Proof:} 

Let $M^*$ be the max-weight matching on $G$.  Since the LP relaxation
is loose, by Prop. \ref{prop:bad_subgr}, there exists either a bad
stemmed blossom or a bad blossom pair w.r.t. $M^*$. Suppose first that
it contains a bad stemmed blossom $B_1$, and consider some $e\in C\cap
M^*$ that is in the ``blossom'' part of $B_1$ (as opposed to the stem)
and also in $M^*$. From the two nodes of $e$, make maximal alternating
paths $P_1$ and $P_2$ that remain in $B_1$ and start out in opposite
directions on $C$. For the stemmed blossom example above, if $e$ is
the edge $(a,b)$ then the two paths will be $bcfghi$ and $aedcfghi$.

Let $ d_1 ~ = ~ w(P_1- M^*) - w(P_1\cap M^*)$, and similarly $d_2$ for
$P_2$. $d_1$ represents the change in the weight of the matching if
each edge in $P_1$ were ``switched'', i.e. their membership in the
matching was reversed from its original value.  It is easy to see that
\begin{eqnarray*}
d_1 + d_2 - w(e) & = & w(C- M^*) + 2w(P- M^*) \\
& & \quad - w(C\cap M^*) - 2w(P\cap M^*) 
\end{eqnarray*}
By assumption $B_1$ is a bad blossom and hence we have that $d_1 + d_2
- w(e) > 0$.

Suppose max-product converges to $M^*$ by step $k$. Consider now the
computation tree $T_e$ which is full upto depth $k+|V|$, where $|V|$
is the number of nodes in $G$. Let $M_T$ be the max-weight matching on
$T_e$. Lemma \ref{lem:convg_tree} implies that $M_T$ will be a
projection of $M^*$ in a distance-$|V|$ neighborhood of the
root. Also, starting from the root $e$, each of $P_1$ and $P_2$ will
have a unique copy, say $R_1$ and $R_2$ respectively, in $T_e$, with
$|R_1|,|R_2| < |V|$. Since $P_1$ and $P_2$ are alternating
w.r.t. $M^*$, it follows that $R_1$ and $R_2$ will be alternating with
respect to $M_T$. Also, the set $S = R_1\cup e \cup R_2$ forms an
alternating path on $T_e$ with respect to $M_T$, and this begins and
ends in nodes unsaturated by $M_T$. Thus, $M_T$ can be augmented by
this path: the set $M_T - (S\cap M_T) + (S-M_T)$ will be a matching on
$T_e$.

Also, the weight gain from doing this augmentation will be exactly
$d_1 + d_2 - w(e)$, which we know is strictly positive. Thus, this
shows that $M_T$ is not the optimal matching on $T_e$, which
contradicts the choice of $M_T$. This means that our assumption about
max-product convergence to $M^*$ is incorrect.

Thus, we see that if there exists a bad stemmed blossom w.r.t. $M^*$
in $G$ then max-product does not converge to $M^*$. A similar argument
holds for the case of a bad blossom pair $B_2$, except that instead of
paths $P_1$ and $P_2$ above we now have to look at alternating walks
$W_1$ and $W_2$ that live in $B_2$ and are long enough. These walks
can then be mapped to an augmenting path on $T_e$ which strictly
improves $M_T$, leading to a contradiction as was seen in the case of
the paths $P_1$ and $P_2$. This completes the proof. \hfill
$\blacksquare$

\section{Discussion}

The results of this paper can be generalized to the case of perfect
matchings, $b$-matchings and perfect $b$-matchings in general graphs,
where similar results hold. In this paper max-product is shown to be
as powerful as LP relaxation, but it would be more interesting to
outline a direct {\em operational} link between max-product and a
linear programming algorithm. As an example, \cite{bib:bayati} shows
that for bipartite matching max-product has an operational
correspondance with the auction algorithm
\cite{bib:bertk_auction}. Also, the form of the message update
equations suggests that it can be implemented via an equivalent
message passing update rule between just the nodes of the graph $G$,
instead of having messages go from nodes to edges and vice versa.

More generally, it would be interesting to see if the ideas presented
in this paper could be used/genealized to show connections between
linear programming and belief propagation in other applications.

\section*{Acknowledgements}

The author would like to acknowledge Dmitry Malioutov, whose
experiments suggested a strong link between LP relaxation and
max-product performance for non-bipartite graphs. Dmitry is also
responsible for pointing the author to the local optimality result
\cite{bib:weiss_local_opt}.

\appendix

\noindent{ \bf Proof of Proposition \ref{prop:bad_subgr}}

We now show that if the LP relaxation is loose then there exists in
the graph either a bad stemmed blossom or a bad blossom pair, with
respect to the optimal matching $M^*$. Let $x$ be the optimal
(fractional) solution to the LP relaxation. 

Let $E'$ be the set of all edges $e$ such that either {\em (a)} $e\in
M^*$, or {\em (b)} $e\notin M^*$ and $x_e>0$.  Then, $E'$ will contain
at least one edge $e\notin M^*$, because if all $e\notin M^*$ had
$x_e=0$ then the LP would be tight.  Let $G'=(V,E')$ be the subgraph
of $G$ having only the edges in $E'$. An {\em cycle augmentation} is
any even cycle in which every alternate edge is in $M^*$. A {\em path
augmentation} is any path in which every alternate edge is in $M^*$,
and which begins and ends in nodes unsaturated by $M^*$. For any
augmentation $A$, we have that $M^* - (A\cap M^*) + (A-M^*)$ is also a
matching in $G'$. Thus, if $M^*$ is the unique max-weight matching it
has to be that $w(A\cap M^*)>w(A-M^*)$.

\begin{lemma}
$G'$ cannot contain any augmentations: cycles or paths.
\end{lemma}

{\em Proof:} Let $A$ be an augmentation in $G'$. By assumption,
$x_e>0$ for all $e\in A-M^*$, which implies that $x_e<1$ for all $e\in
A\cap M^*$. Thus, there exists some $\epsilon >0$ such that decreasing
each $x_e, e\in A-M^*$ by $\epsilon$ and increasing each $x_e,e\in
A\cap M^*$ by $\epsilon$ represents a valid new feasible point for the
LP. The weight of this new point exceeds the weight of $x$ by
$\epsilon (w(A\cap M^*)-w(A-M^*)) > 0$. However this contradicts the
optimality of $x$, and thus $G'$ cannot contain any
augmentation. \hfill $\blacksquare$

Let $S$ be the longest alternating sequence of edges in $G'$, and let
$v_1$ and $v_2$ be its endpoints. By the lemma above, both cannot be
unsaturated. We say that $v_1$ or $v_2$ is a {\em saturated leaf} if
it is saturated by $M^*$ and there exist no edges in $G'-M^*$ incident
on it. Also, note that an endpoint is saturated if and only if its
corresponding edge in $S$ is also in $M^*$.

The fact that $S$ is the longest sequence means that it cannot be
extended further beyond $v_1$ and $v_2$. This implies that one of the
following cases must occur:
\begin{enumerate}
\item Both $v_1$ and $v_2$ are both saturated leaves \\ In this case,
  the constraints at $v_1$ and $v_2$ are loose. So, there exists an
  $\epsilon$ such that if all $x_e, e\in S-M^*$ are decreased by
  $\epsilon$ and all $x_e,e\in S\cap M^*$ are increased by $\epsilon$
  then the new solution remains feasible. This new solution will have
  strictly higher weight than $x$, which is a contradiction. Thus this
  case cannot occur.
\item $v_1$ is a saturated leaf and $v_2$ is unsaturated. \\ An
  $\epsilon$-perturbation argument like the one above can be used to
  show that this case too cannot occur.
\item $v_1$ is saturated by $M^*$. but is not a leaf. $v_2$ is either
  unsaturated, or a saturated leaf. \\ Since $S$ cannot be extended,
  it has to be that all edges in $G'-M^*$ incident to $v_1$ have other
  endpoints in $S$.  Let $e$ be one such edge. Then, $e\cap S$ forms a
  stemmed blossom: the resulting cycle has to be odd, and the
  remaining part of $S$ will be a stem whose endpoint is $v_2$. Note
  that in this case it has to be that the constraint at $v_2$ is
  loose.
\item Both $v_1$ and $v_2$ are saturated by $M^*$, but are not
  leaves. \\ Applying the above blossom argument to both $v_1$ and
  $v_2$ yields the existence of a blossom pair.
\end{enumerate}

Thus if the LP relaxation is loose then there exists a stemmed blossom
or a blossom pair. Now all that is remaining to show is that they are
``bad''. Let $B_1$ be a stemmed blossom in $G'$, consisting of blossom
$C$ and stem $P$. Then, there exists some $\epsilon>0$ such that if
$x_e,e\in C\cap M^*$ is increased by $\epsilon$, $x_e,e\in C-M^*$ is
decreased by $\epsilon$, $x_e,e\in P\cap M^*$ is increased by
$2\epsilon$, and $x_e,e\in C-M^*$ is decreased by $2\epsilon$, then
the new solution remains feasible for the LP. Also, the new solution
weighs 
\[
\epsilon \left [ w(C\cap M^*) + 2w(P\cap M^*) - w(C-M^*) - 2w(P-M^*) \right ]
\]
more than $x$. For $x$ to be the unique optimal of the LP, this has to
be strictly negative and thus any stemmed blossom $B_1$ is bad. A
similar argument shows that any blossom pair is bad. This finishes the
proof of the proposition. \hfill $\blacksquare$


\begin{thebibliography}{10}
\providecommand{\url}[1]{#1}
\csname url@rmstyle\endcsname
\providecommand{\newblock}{\relax}
\providecommand{\bibinfo}[2]{#2}
\providecommand\BIBentrySTDinterwordspacing{\spaceskip=0pt\relax}
\providecommand\BIBentryALTinterwordstretchfactor{4}
\providecommand\BIBentryALTinterwordspacing{\spaceskip=\fontdimen2\font plus
\BIBentryALTinterwordstretchfactor\fontdimen3\font minus
  \fontdimen4\font\relax}
\providecommand\BIBforeignlanguage[2]{{%
\expandafter\ifx\csname l@#1\endcsname\relax
\typeout{** WARNING: IEEEtran.bst: No hyphenation pattern has been}%
\typeout{** loaded for the language `#1'. Using the pattern for}%
\typeout{** the default language instead.}%
\else
\language=\csname l@#1\endcsname
\fi
#2}}

\bibitem{bib:aji_single_cycle}
S.~M. Aji, G.~B. Horn, and R.~J. McEliece, ``On the convergence of iterative
  decoding on graphs with a single cycle,'' in \emph{ISIT}, 1998, p. 276.

\bibitem{bib:weiss_single_cycle}
Y.~Weiss, ``Correctness of local probability propagation in graphical models
  with loops,'' \emph{Neural Computation}, vol.~12, no.~1, pp. 1--41, 2000.

\bibitem{bib:weiss_freeman_gaussian}
Y.~Weiss and W.~Freeman, ``Correctness of belief propagation in gaussian
  graphical models of arbitrary topology,'' \emph{Neural Computation}, vol.~13,
  no.~10, pp. 2173--2200, 2001.

\bibitem{bib:walksums}
D.~Malioutov, J.~Johnson, and A.~Willsky, ``Walk-sums and belief propagation in
  gaussian graphical models,'' \emph{Journal of Machine Learning Research},
  vol.~7, pp. 2031--2064, Oct. 2006.

\bibitem{bib:rich_urb}
T.~Richardson and R.~Urbanke, ``The capacity of low-density parity check codes
  under message-passing decoding,'' \emph{IEEE Transactions on Information
  Theory}, vol.~47, pp. 599--618, 2001.

\bibitem{bib:van_roy}
P.~Rusmevichientong and B.~V. Roy, ``An analysis of belief propagation on the
  turbo decoding graph with gaussian densities,'' \emph{IEEE Transactions on
  Information Theory}, vol.~47, no.~2, pp. 745--765, 2001.

\bibitem{bib:weiss_local_opt}
Y.~Weiss and W.~Freeman, ``On the optimality of solutions of the max-product
  belief-propagation algorithm in arbitrary graphs,'' \emph{IEEE Transactions
  on Information Theory}, vol.~47, no.~2, pp. 736--744, Feb. 2001.

\bibitem{bib:bayati}
M.~Bayati, D.~Shah, and M.~Sharma, ``Maximum weight matching via max-product
  belief propagation,'' in \emph{ISIT}, Sept. 2005, pp. 1763 -- 1767.

\bibitem{bib:b_match}
B.~Huang and T.~Jebara, ``Loopy belief propagation for bipartite maximum weight
  b-matching,'' in \emph{Artificial Intelligence and Statistics (AISTATS)},
  March 2007.

\bibitem{bib:yanover_lp_bp}
C.~Yanover, T.~Meltzer, and Y.~Weiss, ``Linear programming relaxations and
  belief propagation -- an empirical study,'' \emph{Jourmal of Machine Learning
  Research}, vol.~7, pp. 1887--1907, 2006.

\bibitem{bib:wainw_linprog}
M.~Wainwright, T.~Jaakkola, and A.~Willsky, ``Map estimation via agreement on
  (hyper)trees: Message-passing and linear-programming approaches,'' \emph{IEEE
  Transactions on Information Theory}, vol.~51, no.~11, pp. 3697--3717, Nov.
  2005.

\bibitem{bib:feldman_allerton}
J.~Feldman, D.~Karger, and M.~Wainwright, ``Linear programming-based decoding
  of turbo-like codes and its relation to iterative approaches.'' in
  \emph{Allerton Conference on Communication, Control, and Computing}, 2002.

\bibitem{bib:lp_decoding}
J.~Feldman, M.~Wainwright, and D.~Karger, ``Using linear programming to decode
  binary linear codes,'' \emph{IEEE Transactions on Information Theory},
  vol.~51, pp. 954--972, 2005.

\bibitem{bib:ksh_frey_loel}
F.~Kschischang, B.~Frey, and H.~Loeliger, ``Factor graphs and the sum-product
  algorithm,'' \emph{IEEE Transactions on Information Theory}, vol.~47, no.~2,
  pp. 498--519, Feb. 2001.

\bibitem{bib:tatik_jordan}
S.~Tatikonda and M.~Jordan, ``Loopy belief propagation and gibbs measures,'' in
  \emph{Uncertainty in Artificial Intelligence}, vol.~18, 2002, pp. 493--500.

\bibitem{bib:edmonds}
J.~Edmonds, ``Paths, trees and flowers,'' \emph{Canadian Journal of
  Mathematics}, vol.~17, pp. 449--467, 1965.

\bibitem{bib:bertk_auction}
D.~Bertsekas, ``Auction algorithms for network flow problems: A tutorial
  introduction,'' \emph{Computational Optimization and Applications}, vol.~1,
  pp. 7--66, 1992.

\end{thebibliography}


\end{document}